
\magnification=1200
{\nopagenumbers
\baselineskip=15pt
\hsize=5.5in

\hskip10cm UM--P-94/25\hfil\par
\hskip10cm OZ--94/12\hfil\par
\hskip10cm PITHA 94/17\hfil\par
\vskip20pt
\centerline{\bf Transverse Polarization of Top Quarks} \par
\centerline{\bf Produced at a Photon-Photon Collider}\par
\vskip1.5cm
\hskip4cm W. Bernreuther\par
\vskip0.5cm
\hskip4cm Institut f\"ur Theoretische Physik  \par
\hskip4cm Physikzentrum RWTH Aachen \par
\hskip4cm 52056 Aachen, Germany \par
\vskip0.5cm
\hskip4cm J.P. Ma and B.H.J. McKellar \par
\vskip0.5cm
\hskip4cm Research Center for High Energy Physics \par
\hskip4cm School of Physics\par
\hskip4cm University of Melbourne \par
\hskip4cm Parkville, Victoria 3052\par
\hskip4cm Australia \par
\vskip1cm
{\bf\underbar{Abstract}}:\par
At future $\gamma \gamma$ colliders copious production of $t \bar{t}$ pairs
is possible. This would allow a detailed investigation of the
interactions involving the top quark. We propose some orrelations which
are sensitive to $t \bar{t}$ final state interactions and we compute
the QCD and standard model Higgs boson contributions to these correlations.
QCD induced transverse polarization of top quarks is found to be sizeable
and measurable at a high-energy $e^+e^-$ collider with an integrated
luminosity of 10 $({\rm fb})^{-1}$ which is converted into a photon collider
by backscattering of laser photons.
\par\vfil\eject }
\baselineskip=20pt
\pageno=1
One of the attractive possibilities of a future high energy linear
$e^+e^-$ collider [1] is to convert it via backscattering of
laser photons off the initial lepton beams into a high energy $\gamma
\gamma$ collider [2]. It is expected that such a facility would
provide a tool for a number of novel precision studies of strong and
electroweak interactions [1, 3 - 8]. For instance copious
production of top quarks
is feasible. This would allow for some detailed studies of
top quark physics [5 - 8] which would be to some extent complementary
\footnote*{For $t \bar{t}$ production by beamstrahlung photons at a
linear $e^+e^-$ collider see [9,10].}
to studies of $e^+e^- \rightarrow t \bar{t}$.
One important aspect of top quark physics is the quasi-free behaviour
of the top due to its heavy mass ( $m_t > 130$GeV [11]). On average the
top will have decayed before being able to form hadrons. This property
makes the the spin polarization of the top a good observable as it
can be traced through the angular distributions of the $t$ and/or
$\bar{t}$ decay products.
\par
In this note we exploit this property and investigate the transverse
polarization (i.e., the polarization transverse to the production plane)
of the $t$ and/or $\bar{t}$ produced in $\gamma \gamma$ collisions.
This polarization which is due to  $t \bar{t}$ final state
interactions may serve as a probe of (non) standard model (SM)
interactions [12 - 15] in the $t \bar{t}$ system. We propose
observables by which this tranverse polarization can be traced
in the $t$ and $\bar{t}$ decay products
and compute the dominant  SM contributions to their expectation
values.

\par
We consider $t\bar t$ production via photon fusion:
 $$ \gamma (p_1)+\gamma (p_2) \rightarrow t(k_1, s_1) +\bar t(k_2,s_2)
    \eqno(1) $$
where the momenta are defined in the photon-photon
c. m. frame and $s_1$ and $s_2$ label the
spins of $t$ and $\bar{t}$.
In the following we consider only unpolarized photon beams.
The process may then be described by the density matrix:
  $$ R_{\alpha \alpha ',\beta \beta '} =\sum\   '
  < t(k_1,\alpha ')\bar t(k_2,\beta ') \vert T \vert \gamma\gamma >^*
  < t(k_1,\alpha )\bar t(k_2,\beta ) \vert T \vert \gamma\gamma >
   \eqno (2) $$
where the $\sum '$ denotes averaging over the $\gamma \gamma$ polarizations.
Note that R is an even function of the three-momentum
${\bf p_1}$ due to Bose symmetry of the
two-photon state. With respect to the $ t \bar{t}$
spin space it has the following matrix structure:
  $$ R= A I\times I +{\bf B}\cdot \sigma \times I
           +{\bf C} \cdot I\times \sigma +D_{ij}\sigma^i\times
    \sigma^j \eqno(3) $$
where $\sigma^i$ denote the Pauli matrices and the first (second) factor
in the tensor products refers to the $t (\bar t)$ spin space.
Due to rotation invariance
the structure functions ${\bf B}$, ${\bf C}$ and $D_{ij}$ can be
decomposed in terms of the unit vectors
 ${\bf \hat p_1}$, ${\bf \hat k_1}$ and ${\bf \hat n}=
 {\bf \hat p_1}\times {\bf \hat k_1} /
  \vert {\bf \hat p_1}\times {\bf \hat k_1}\vert $ which
is orthogonal to the production plane. For ${\bf B}$ and  ${\bf C}$
 one can write:
 $$ {\bf B}=b_1{\bf \hat p_1}+b_2 {\bf \hat k_1}+b_3 {\bf \hat n}, \ \
   {\bf C}=c_1{\bf \hat p_1}+c_2 {\bf \hat k_1}+c_3 {\bf \hat n}, \ \
 \eqno(4) $$
while $D_{ij}$ which characterizes the correlation between the $t$ and
$ \bar t$ spins can be decomposed in terms of 8
independent scalar functions. For the tree-level SM amplitude one has
 ${\bf B}={\bf C}={\bf 0}$, i.e., the top quarks are not polarized.
If time reversal (T)
invariance holds the structure functions $b_3$ and $c_3$, which lead to
$t$ and $\bar{t}$ polarization transverse to the production
plane, become nonzero  only
due to the interference between the dispersive part and the absorptive
part of the scattering amplitude for the process (1).
It is easy to show that due to
Bose symmetry of the $\gamma \gamma$ state the relations
     $$  b_3(x) = -b_3(-x), \ \ c_3(x) = - c_3(-x) \eqno(5) $$
must hold, where $x = {\bf \hat p_1}\cdot {\bf \hat k_1}$ is the cosine of the
scattering angle. This implies that at the level of $t \bar{t}$ final states
the spin projection ${\bf s_1}\cdot {\bf \hat n}$, where  ${\bf s_1}={1\over 2}
  \sigma \times I$, must be weighted with $x$ in order to have a
non-zero expectation value, and likewise for the
corresponding $\bar{t}$ spin projection. Furthermore,
one may wish to disentangle
final-state interaction effects from CP-violating phenomena, which may also
occur in the $t \bar{t}$ system. For detecting the former effects it is
useful to employ T-odd (i.e., odd under reflection of momenta and spins)
but CP-even correlations [14]. Hence for studying
final state interactions  in the process (1) which are induced by
CP-invariant interactions the appropriate observable would be
 $$ \eqalign { x ({\bf s_1} +{\bf s_2})\cdot {\bf \hat{n}}
 = {x\over 2} (\sigma \times
  I + I\times \sigma)\cdot {\bf \hat{n}}} \eqno(6)$$
Of course this cannot be used in an experiment
because measurements of the $t$ and $\bar{t}$ spin cannot be made
on an event-by-event basis.
\par
As mentioned above the top quark analyzes its spin by its parity-violating
weak decay $t\rightarrow W + b$. It is well-known that the charged lepton
from subsequent $W$ decay is an efficient spin analyzer of the $t$ quark [16].
In order to construct realistic observables we consider
first the case where both $t$ and $\bar{t}$ decay semileptonically, i.e.,
  $$ \eqalign {t &\rightarrow W^+ +b \rightarrow \ell ^+ +\nu_{\ell} +b \cr
            \bar t & \rightarrow W^- +\bar b \rightarrow
      \ell ^- +{\bar\nu_{\ell}} +\bar b \cr} \eqno(7)$$
A CP-even but T-odd observable can then be formed using the momenta
${\bf q_{\pm}}$ of the charged leptons $\ell ^{\pm}$ measured in the
$e^+e^-$ laboratory frame and the direction ${\bf \hat p}$
of the electron beam:
  $$ O_L={1\over m_t^3} \{ {\bf \hat p} \cdot ({\bf q_+} +{\bf q_-})\}
          {\bf \hat p}\cdot ({\bf q_+}\times {\bf q_-}) \eqno(8) $$
where the top mass $m_t$ is used to make $O_L$ dimensionless.
\par
The other useful type of events results from semileptonic $t$ and
non-leptonic $\bar{t}$ decay and vice versa.
(These will be called "semihadronic" below.) For these events the
momentum direction  ${\bf \hat k_-} ({\bf \hat k_+})$
of the $\bar t (t)$ can be reconstructed. (Again we refer to the laboratory
frame.) For the respective channels we can then use the T-odd observables:

  $$ \eqalign { O_{B_1} &= {1\over m_t}  ({\bf \hat p}\cdot {\bf\hat k_-})
                  {\bf\hat p}\cdot ({\bf\hat k_-}\times {\bf q_+}) \cr
 O_{B_2} &= {1\over m_t}  ({\bf \hat p}\cdot {\bf\hat k_+})
                  {\bf\hat p}\cdot ({\bf\hat k_+}\times {\bf q_-}) . \cr}
  \eqno(9) $$
As these observables are intimately related to
the $t$ and $\bar{t}$
spin-momentum projections $x{\bf s_{1,2}}\cdot {\bf \hat n}$,
they are more sensitive to $t \bar{t}$
final-state interaction effects than (8).
The translation of the spin-momentum correlation (6) into a final state
observable is $O_{B_1} - O_{B_2}$ which is CP-even. Below we shall evaluate:
$$ \eqalign{B_t &=  <O_{B_1}> - <O_{B_2}>.} \eqno(10)$$
In this quantity contributions from CP-invariant absorptive parts add up.
\par
Within the SM the dominant contributions to the absorptive part of the
scattering amplitude of (1) arise from QCD corrections and possibly
also from Higgs boson ($H$) exchange due to the sizeable
Yukawa coupling of the top quark. Of the QCD corrections, only gluon exchange
between the final $t$ and $\bar{t}$ quarks leads to an absorptive part
of the one-loop amplitude. An absorptive part from Higgs boson interactions
arises from $H$ exchange between the final $t$ and $\bar{t}$ and from
s-channel $H$ exchange with a $\gamma \gamma H$ vertex being induced by
$W$ boson and $t$ quark loops. However, interference
of the s-channel diagrams with the Born amplitude does not induce
a transverse polarization of the top quark.
\par
In general, the interference between the Born amplitude and
the absorptive part will give contributions not only to $b_3$ and
$c_3$ but also to some terms in $D_{ij}$. However, in the case at hand,
only $b_3$ and $c_3$ become nonzero.  Gluon exchange gives:
  $$\eqalign { b^{gluon}_3 = c^{gluon}_3 &= \alpha_s e^4 Q_t^4
|{\bf \hat p_1}\times {\bf \hat k_1}| (m_t/\beta E_1)\cdot (1-\beta^2 x^2)^{-1}
\cdot (x^2 - 1)^{-1} \cr
      & \cdot [4\beta x (1-x^2)
+2\beta x (-5\beta^2-6\beta - 1) \ln (\beta +\beta^2)\cr
&+2\beta x(5\beta^2 - 6\beta +1) \ln(\beta - \beta^2) \cr
& +2\beta (-2\beta^2 x^2 - 3\beta^2 +6\beta x - x^2) \ln(\beta - \beta^2 x) \cr
& +2\beta (2\beta^2 x^2 + 3\beta^2 +6\beta x + x^2) \ln(\beta + \beta^2 x) ]
  \cr }
 \eqno(11)$$
where $Q_t$ is the charge of the top quark in units of $e$, $E_1$ is the
photon energy, $\beta = (1 - m_t^2/E_1^2)^{1/2}$
is the velocity of the top quark
and $x$ is the cosine of the scattering angle defined above.
Higgs boson exchange induces:
  $$\eqalign { b^{Higgs}_3 &= c^{Higgs}_3   ={3e^4Q_t^4 \over 32\pi}
        {m_t^2\over v^2}
                 |{\bf \hat p_1}\times {\bf \hat k_1}| (m_t/E_1)
      (1-\beta^2 x^2)^{-1}(x^2-1)^{-1}
     \cr
      &\cdot \{ 2\beta^2x(x^2+2d_H-3){\rm ln} {1+\beta\over 1-\beta}
           +2\beta x(x^2d_H-3d_H+2) {\rm ln }{d_H+1 \over d_H-1} \cr
      & +\beta x(-\beta^2x^2d_H-\beta^2x^2-2\beta^2d_H^2 +3\beta^2d_H
     +\beta^2 +2d_H-3) (F(\beta)+F(-\beta) ) \cr
      & +(\beta^2 x^4+3\beta^2x^2d_H -2\beta^2x^2-2\beta^2d_H^2 +\beta^2 d_H
      -\beta^2 -x^2 +1)(F(\beta)-F(-\beta) ) \cr
      &   +4\beta x(1-x^2) \}, \ \ \
      d_H =1+{M_H^2 \over 2E_1^2\beta^2 }, \cr
   &  F(\beta) =
      {1\over R} {\rm ln}{d_H-\beta x+R \over d_H-\beta x -R }, \ \
     R=\sqrt{ (1-\beta x d_H)^2 +\beta^2 (d_H^2-1)(1-x^2) } \cr
     }  \eqno(12) $$
where $v = $246 GeV.
Using in addition the contribution of the Born ampitude to $R$, which is
easily calculated, we can now compute the expectation values of (8) and (9).
As further ingredients for this calculation one needs
the distributions
of polarized $t$ and $\bar{t}$ decay into the respective channels. We take the
distributions as obtained from the SM Born decay amplitudes. Finally one
requires the $\gamma \gamma$ luminosity spectrum which we take from [2].
We assume the energy of the laser
photon to be 1.26eV and the $e-\gamma$ conversion
factor is taken to be one. Then, choosing two different values for the
top mass,
we obtain  for a $e^+e^-$ collider at center-of-mass energy
$\sqrt{s} =500$GeV :
  $$\eqalign { <O_L> & = 0.0037\alpha_s ,\ \
               B_t = 0.11\alpha_s ,
       \ \ {\rm for}\ m_t=130{\rm GeV}, \cr
       <O_L> & = 3.9\times 10^{-4}\alpha_s ,\ \
               B_t =0.047 \alpha_s ,
       \ \ {\rm for}\ m_t=170{\rm GeV}. \cr }\eqno (13) $$
For $\sqrt{s}=1$TeV we get:
   $$\eqalign { <O_L> & = 0.041\alpha_s ,\ \
               B_t = 0.23\alpha_s ,
       \ \ {\rm for}\ m_t=130{\rm GeV} , \cr
       <O_L> & = 0.016\alpha_s ,\ \
               B_t = 0.18\alpha_s ,
       \ \ {\rm for}\ m_t=170{\rm GeV} .\cr }\eqno (14)$$
 Eqs. (13) and (14) do not contain the contributions from Higgs boson
exchange. They are very small and of the order of $10^{-5}$ to $10^{-8}$.
In order to estimate the sensitivity of the
correlations to the QCD induced
transverse polarization we need the effective $t \bar{t}$
cross sections (i.e., the cross sections for the
process (1) folded with the $\gamma \gamma$ luminosity spectrum)
and the width of the
distribution of the observables  $\Delta O =
(<O^2> - <O>^2)^{1/2} \simeq \hfill\break
(<O^2>)^{1/2}$. For $m_t=130$GeV:
  $$ \eqalign { \sigma_{t \bar{t}}=& 0.52{\rm pb}, \ \
      <O_L^2> =0.0016, \ \ <O_{B_1}^2 >=0.0065 , \ {\rm
 at }\ \sqrt{s}=500{\rm GeV}, \cr
   \sigma_{t \bar{t}}=& 0.995{\rm pb}, \ \
      <O_L^2> =0.038\ \ <O_{B_1}^2 >=0.005 , \ {\rm
 at }\ \sqrt{s}=1000{\rm GeV}. \cr} \eqno(15) $$
and for $m_t=170$GeV:
  $$ \eqalign { \sigma_{t \bar{t}}=& 0.09{\rm pb}, \ \
      <O_L^2> =0.00026 \ \ <O_{B_1}^2 >=0.0056 , \ {\rm
 at }\ \sqrt{s}=500{\rm GeV}, \cr
   \sigma_{t \bar{t}}=& 0.60{\rm pb}, \ \
      <O_L^2> =0.0092\ \ <O_{B_1}^2 >=0.0044 , \ {\rm
 at }\ \sqrt{s}=1000{\rm GeV}. \cr} \eqno(16) $$
The numbers in eqs. (13),(14) and (15),(16) show that the sensitivity
-- signified by the signal-to-noise ratios
$<O>/\Delta O$ --  of the observables
(9) is much higher than this of the one obtainable with (8).
For $m_t$ = 130 GeV we have
$B_t/(\sqrt{2}\Delta O_{B_1})
 \simeq 0.10(0.23)$  at $\sqrt{s} = 500 (1000)$ GeV.
In order to establish this correlation as a, say, 4$\sigma$ effect about
1600 (320) "semihadronic" $t \bar{t}$ events are required at the respective
energies, which is feasible with an luminosity of
10 $({\rm fb})^{-1}$ for $e^+e^-$ colliders.
For $m_t$ = 170 GeV the corresponding numbers are
$B_t/(\sqrt{2}\Delta O_{B_1})
 \simeq 0.044(0.2)$  at $\sqrt{s} = 500 (1000)$ GeV.
Using the cross sections of (16) one sees that in this case one
obtains only a 1$\sigma$ effect at 500 GeV, whereas the correlation (9)
would be clearly measurable (as a 10$\sigma$ effect with the above
luminosity) at 1 TeV.
\par
In conclusion: we have shown that the triple product correlations (9) are
sensitive tools for detecting transverse polarization of $t$ and $\bar{t}$
produced in $\gamma \gamma$ collisions. Transverse polarization
of the top quark is due to
radiative corrections and results within the SM primarily from QCD final
state interactions. As the couplings of the top quark have yet to be
measured in future experiments the observables above may be used for
a detailed investigation of the forces in the $t \bar{t}$ system.

\vfil\eject

{\bf REFERENCES}
\par
[1] "$e^+ e^-$ Collisions at 500 GeV: The Physics Potential",
ed. by P.M. Zerwas,
DESY publication DESY 92-123A,B, Hamburg 1992;
"Physics and Experiments \par \noindent with Linear Colliders", ed. by R.
Orava,
P. Eerola, and
M. Nordberg (World Scientific, Singapore) , Vols. I, II (1992).\hfill

[2] I.F. Ginzburg et al., Nucl. Instrum. Meth. 205 (1983) 47;
ibid. 219 (1984) 5; V.I. Telnov, Nucl. Instr. Meth. 294 (1990) 72.\hfill

[3] S.Y. Choi and F. Schrempp, Phys. Lett. B227 (1991) 149.\hfill

[4] H. Veltman, Saclay preprint S. PH.T93/111 (1993).\hfill

[5] E. Boos et al., Z. Phys. C56 (1992) 487.\hfill

[6] J.H. K\"uhn, E. Mirkes, and J. Steegborn, Z. Phys. C57 (1993) 615.\hfill

[7] O.J.P. Eboli et al., Madison preprint MAD-PH-701 (1993).\hfill

[8] I.I. Bigi, F. Gabbiani, and V.A. Khoze, Nucl. Phys. B406 (1993) 3.\hfill

[9] F. Halzen, C.S. Kim, and M.L. Stong, Phys. Lett. B274 (1992) 489.\hfill

[10] M. Drees, M. Kr\"amer, J. Zunft, and P.M. Zerwas, Phys. Lett.
B306 (1993) 371. \hfill

[11] S. Abachi et al. (D0 Collaboration), FERMILAB Pub-94/004-E (1994)\hfill

[12]J.H. K\"uhn, A. Reiter, and P.M. Zerwas, Nucl. Phys. B272 (1986) 560.\hfill

[13] G. L. Kane, G.A. Ladinski, and C.-P. Yuan, Phys. Rev. D45 (1992)
124.\hfill

[14] W. Bernreuther, J.P. Ma, and T. Schr\"oder, Phys. Lett B297 (1992)
218\hfil

[15] T. Arens and L.M. Sehgal, Nucl. Phys. B393 (1993) 46\hfill

[16] A. Czarnecki, M. Jezabek, and J.H. K\"uhn, Nucl. Phys. B351 (1991) 70.
\vfil\eject\end